\documentclass[journal]{IEEEtran}

\ifCLASSINFOpdf
\else
\fi

\usepackage{framed,multirow}
\usepackage{graphicx}
\usepackage{amssymb}
\usepackage{amsmath}
\usepackage{multirow}
\usepackage{booktabs}
\usepackage{array}
\usepackage{todonotes}
\newcolumntype{L}[1]{>{\raggedright\let\newline\\\arraybackslash\hspace{0pt}}m{#1}}
\newcolumntype{C}[1]{>{\centering\let\newline\\\arraybackslash\hspace{0pt}}m{#1}}
\newcolumntype{R}[1]{>{\raggedleft\let\newline\\\arraybackslash\hspace{0pt}}m{#1}}

\usepackage{bm}

\usepackage{lipsum}
\usepackage{changepage}
\usepackage{color,soul}
\usepackage{amssymb}
\usepackage{amsmath}
\usepackage{enumitem}   

\usepackage{float}

\usepackage{booktabs,xcolor}
\definecolor{lightgray}{gray}{0.9}

\usepackage{tikz}

\usepackage{lipsum}

\let\OLDthebibliography\thebibliography
\renewcommand\thebibliography[1]{
  \OLDthebibliography{#1}
  \setlength{\parskip}{0pt}
  \setlength{\itemsep}{0pt plus 0.3ex}
}

\usepackage{array}

\makeatletter
\newcommand{\thickhline}{%
    \noalign {\ifnum 0=`}\fi \hrule height 1pt
    \futurelet \reserved@a \@xhline
}
\newcolumntype{"}{@{\hskip\tabcolsep\vrule width 1pt\hskip\tabcolsep}}
\makeatother

\usepackage{amssymb}
\usepackage{latexsym}

\usepackage{url}
\usepackage{xcolor}

\usepackage{hyperref}

\definecolor{newcolor}{rgb}{.8,.349,.1}

\begin{document}

\onecolumn
\begin{center}
\Huge{Convolution-Free Medical Image Segmentation \\ using Transformers}
\end{center}

\begin{center}
\normalsize
Davood Karimi, Serge Vasylechko , and Ali Gholipour  \\ Department of Radiology, Boston Children’s Hospital, Harvard Medical School, Boston, MA, USA \end{center}

\vspace{2mm}

\begin{abstract}

Like other applications in computer vision, medical image segmentation has been most successfully addressed using deep learning models that rely on the convolution operation as their main building block. Convolutions enjoy important properties such as sparse interactions, weight sharing, and translation equivariance. These properties give convolutional neural networks (CNNs) a strong and useful inductive bias for vision tasks. However, recent works have also highlighted the limitations of CNNs for segmenting fine and complex structures in medical images. In this work we show that a different deep neural network architecture, based entirely on self-attention between neighboring image patches and without any convolution operations, can achieve more accurate segmentations than CNNs. Given a 3D image block, our network divides it into $n^3$ 3D patches, where $n=3 \text{ or } 5$ and computes a 1D embedding for each patch. The network predicts the segmentation map for the center patch of the block based on the self-attention between these patch embeddings. We show that the proposed model can achieve segmentation accuracies that are better than the state of the art CNNs on three datasets. We also propose methods for pre-training this model on large corpora of unlabeled images. Our experiments show that with pre-training the advantage of our proposed network over CNNs can be significant when labeled training data is small.

\footnotesize
\hspace{5mm} \textbf{Index Terms: }  Medical image segmentation, deep learning, transformers, attention

\end{abstract}

\begin{figure*}[hbt!]
\begin{minipage}[b]{0.9\linewidth}
  \centering
  \centerline{\includegraphics[width=\textwidth]{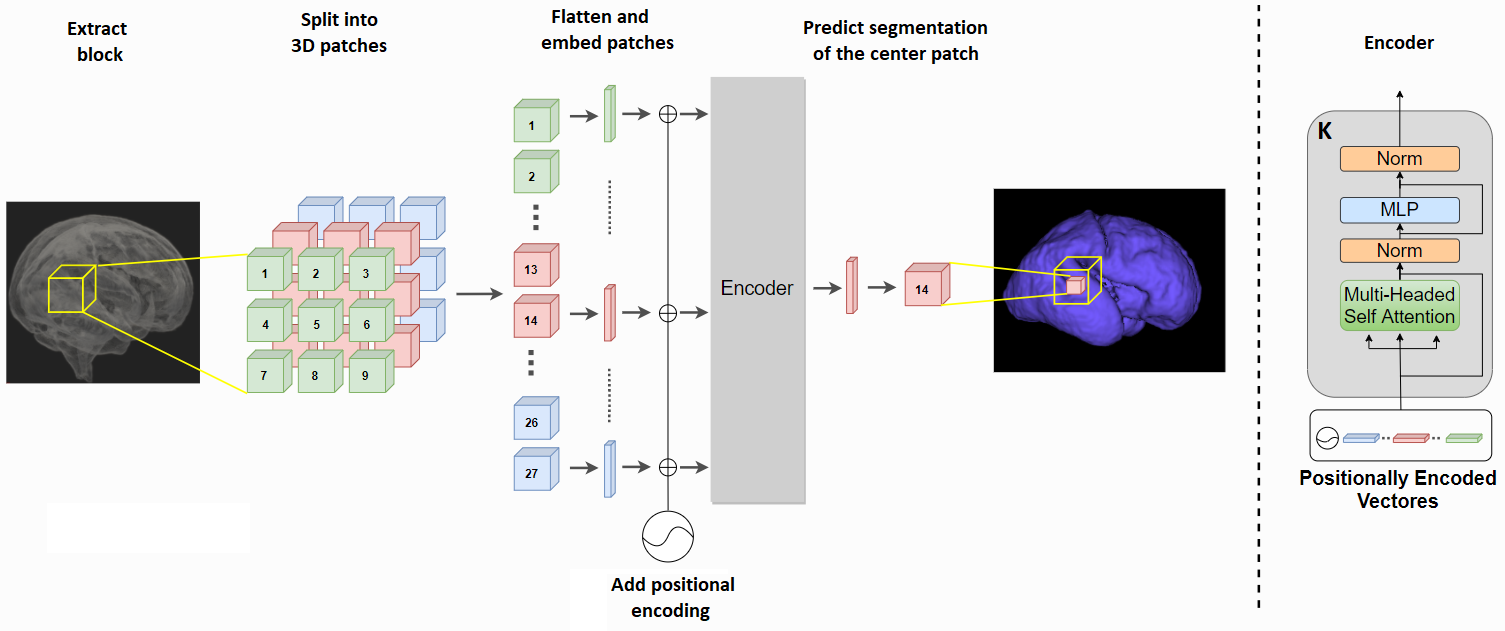}}
  \end{minipage}
\caption{Proposed convolution-free network for 3D medical image segmentation.} \label{network}\end{figure*}

\begin{figure*}[hbt!]
\begin{minipage}[b]{0.9\linewidth}
  \centering
  \centerline{\includegraphics[width=12.0cm]{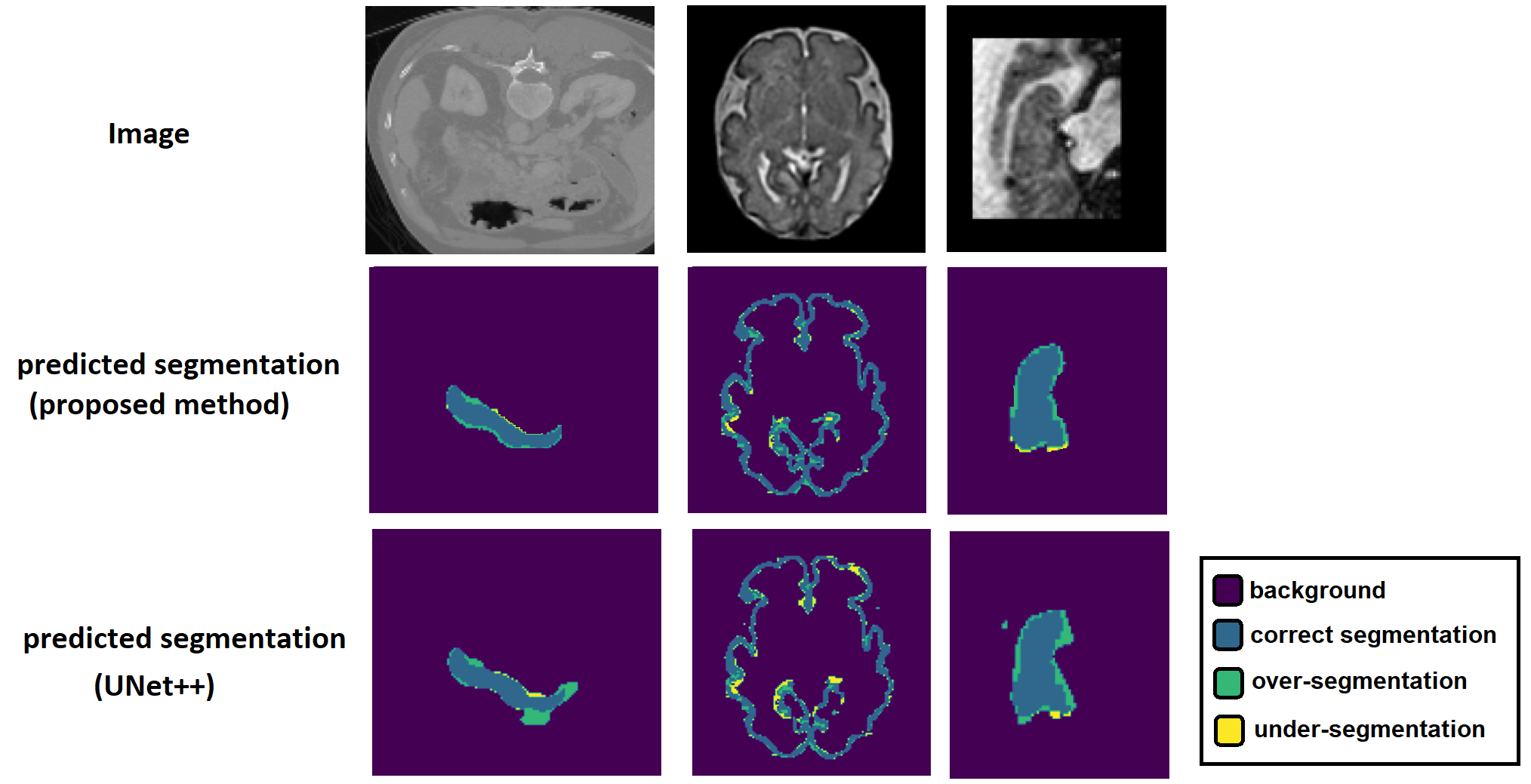}}
  \end{minipage}
\caption{Example segmentations predicted by the proposed network and a state of the art CNN.} \label{res}\end{figure*}

\normalsize
\twocolumn

\section{Introduction}

Image segmentation is a central task in medical image analysis. It is routinely used for quantifying the size and shape of the volume/organ of interest, population studies, disease quantification, treatment planning, and computer-aided intervention. In most medical applications, manual segmentation by an expert is regarded as the gold standard. However, automatic methods promise much faster, cheaper, and more reproducible segmentations.

Classical methods in medical image segmentation run the gamut from region growing \cite{gibbs1996tumour} and deformable models \cite{wang2007medical} to atlas-based methods \cite{thompson1997detection}, Bayesian approaches \cite{prince1995optimization}, graph cuts \cite{mahapatra2014}, clustering methods \cite{goldszal1998image}, and more. In the past few years, deep learning (DL) methods have emerged as a highly competitive alternative and they have achieved remarkable levels of accuracy \cite{bakas2018identifying} \cite{bernard2018deep} \cite{kamnitsas2017efficient} \cite{karimi2019accurate} \cite{zeng2018prostate} \cite{zeng2019liver} \cite{karimi2018segmentation}. It appears that DL methods have largely replaced the classical methods for medical image segmentation.

Due to their success, DL methods for medical image segmentation have attracted much attention from the research community. Recent surveys of the published research on this topic can be found in \cite{hesamian2019deep} \cite{taghanaki2020deep}. Some of the main directions of these studies include improving network architectures, loss functions, and training strategies. Surprisingly, the one common feature in all of these works is the use of the convolution operation as the main building block of the networks. The proposed network architectures differ in terms of the way the convolutional operations are arranged, but they all build upon the same convolution operation. There have been several attempts to use recurrent networks \cite{gao2018} \cite{bai2018recurrent} and attention mechanisms \cite{chen2021transunet} for medical image segmentation. However, all of those models still rely on the convolution operation. Some recent studies have gone so far as to suggest that in practice most of these architectures achieve largely similar results \cite{isensee2018}.

Similar to semantic segmentation, other computer vision tasks such as image classification and object detection have also been most successfully addressed with convolutional neural networks (CNNs) \cite{ren2015}, \cite{krizhevsky2012}. These observations attest to the importance of the convolution operation to image understanding and modeling. The effectiveness of the convolution operation can be attributed to a number of key properties, including: 1) local (sparse) connections, 2) parameter (weight) sharing, and 3) translation equivariance \cite{lecun2015} \cite{le1989handwritten}. These properties are in large part inspired by the neuroscience of vision \cite{olshausen1996}, and give CNNs a very strong inductive bias. In fact, a convolutional layer can be interpreted as a fully connected layer with an ``inﬁnitely strong prior" over its weights \cite{goodfellow2016}. As a result, modern deep CNNs have been able to tackle a variety of vision tasks with amazing success.

In other important applications, most prominently in natural language processing, the dominant architectures were those based on recurrent neural networks (RNNs) \cite{chung2014empirical}, \cite{hochreiter1997long}. The dominance of RNNs ended with the introduction of the transformer model, which proposed to replace the recurrence with an attention mechanism that could learn the relationship between different parts of a sequence as a whole \cite{vaswani2017attention}.

The attention mechanism has had a profound impact on the field of natural language processing. In vision applications, on the other hand, its impact has been much more limited. A recent survey of applications of transformer networks in computer vision can be found in \cite{khan2021transformers}. The number of pixels in typical images is much larger than the number of units of data (e.g., words) in natural language processing applications. This makes it impossible to apply standard attention models to images. As a result, despite numerous efforts (e.g., \cite{wang2020axial} \cite{ho2019axial} \cite{dou2020deep}) attention mechanism has not yet resulted in the type of sea change that has occurred in natural language processing. The recently-proposed vision transformer (ViT) appears to be a major step towards adapting transformer/attention models for computer vision applications \cite{dosovitskiy2020image}. The main insight in that work is to consider image patches, rather than pixels, to be the units of information in images. ViT embeds image patches into a shared space and learns the relation between these embeddings using self-attention modules. Given massive amounts of training data and computational resources, ViT was shown to surpass CNNs in image classification accuracy. Subsequent work has shown that by using knowledge distillation from a CNN teacher and using standard training data and computational resources, transformer networks can achieve image classification accuracy levels on par with CNNs \cite{touvron2020training}.

The goal of this work is to explore the potential of self attention-based deep neural networks for 3D medical image segmentation. We propose a network architecture that is based on self-attention between linear embeddings of 3D image patches, without any convolution operations. We train the network on three medical image segmentation datasets and compare its performance with a state of the art CNN. The contributions of this work are as follows:

\begin{enumerate}

\item We propose the first convolution-free deep neural network for medical image segmentation.

\item We show that our proposed network can achieve segmentation accuracies that are better than or at least on par with a state of the art CNN on three different medical image segmentation datasets. We show that, unlike recent works on image classification (\cite{dosovitskiy2020image} \cite{touvron2020training}), our network can be effectively trained for 3D medical image segmentation with datasets of $\sim 20-200$ labeled images.

\item We propose pre-training methods that can improve our network's segmentation accuracy when large corpora of unlabeled training images are available. We show that when labeled training images are fewer in number, our network performs better than a state of the art CNN with pre-training.

\end{enumerate}

\section{Materials and methods}

\subsection{Proposed network}

Figure \ref{network} shows our proposed network for convolution-free medical image segmentation. The input to the network is a 3D block $B \in {\rm I\!R}^{W \times W \times W \times c}$, where $W$ denotes the extent of the block (in voxels) in each dimension and $c$ is the number of image channels. The block $B$ is partitioned into $n^3$ contiguous non-overlapping 3D patches $\{ p_i \in {\rm I\!R}^{w \times w \times w \times c}  \}_{i=1}^{N}$, where $w= W/n$ is the side length of each patch and $N= n^3$ denotes the number of patches in the block. We choose $n$ to be an odd number. In the experiments presented in this paper $n$ is either 3 or 5, corresponding to 27 or 125 patches, respectively. The network uses the image information in all $N$ patches in $B$ to predict the segmentation for the center patch, as described below.

Each of the $N$ patches is flattened into a vector of size ${\rm I\!R}^{w^3c}$ and embedded into ${\rm I\!R}^D$ using a trainable mapping $E \in {\rm I\!R}^{D \times w^3c}$. Unlike \cite{dosovitskiy2020image}, we do not use any additional tokens since that did not improve the segmentation performance of our network in any way. The sequence of embedded patches $X^0= [E p_1; ...; E p_N] + E_{\text{pos}}$ constitutes the input to our network. The matrix $E_{\text{pos}} \in {\rm I\!R}^{D \times N}$ is intended to learn a positional encoding. Without such positional information, the transformer ignores the ordering of the input sequence because the attention mechanism is permutation-invariant. Therefore, in most NLP applications such embedding has been very important. For image classification, on the other hand, authors of \cite{dosovitskiy2020image} found that positional encoding resulted in relatively small improvements in accuracy and that simple 1D raster encoding was as good as more elaborate 2D positional encodings. For the experiments presented in this paper, we left the positional encoding as a free parameter to be learned along with the network parameters during training because we do not know a priori what type of positional encoding would be useful. We discuss the results of experimental comparisons with different positional encodings below.

The encoder has $K$ stages, each consisting of a multi-head self-attention (MSA) and a subsequent two-layer fully connected feed-forward network (FFN). Both the MSA and FFN modules include residual connections, ReLU activations, and layer normalization. More specifically, starting with the sequence of embedded and position-encoded patches, $X^0$ described above, the $k^{\text{th}}$ stage of the encoder will perform the following operations to map $X^k$ to $X^{k+1}$:

\begin{enumerate}

\item $X^k$ goes through $n_h$ separate heads in MSA. The $i^{\text{th}}$ head:

\begin{enumerate}

\item Computes the query, key, and value sequences from the input sequence:

$$Q^{k,i}= E_Q^{k,i} X^k, K^{k,i}= E_K^{k,i} X^k, V^{k,i}= E_V^{k,i} X^k$$ 

$$\text{where } E_Q, E_K, E_v \in {\rm I\!R}^{D_h \times D}$$

\item Computes the self-attention matrix and then the transformed values:

$$A^{k,i}= \text{Softmax} (Q^TK)/\sqrt{D_h}$$

$$\text{SA}^{k,i}= A^{k,i} V^{k,i}$$

\end{enumerate}

\item Outputs of the $n_h$ heads are stacked and reprojected back onto ${\rm I\!R}^{D}$

$$\text{MSA}^{k}= E_{\text{reproj}}^{k} [\text{SA}^{k,0}; ...; \text{SA}^{k,n_h}]^T$$

$$ \text{where } E_{\text{reproj}} \in {\rm I\!R}^{D \times D_h n_h}$$

\item The output of the MSA module is computed as:
$$X_{\text{MSA}}^{k}= \text{MSA}^{k} + X^k$$

\item $X_{\text{MSA}}^{k}$ goes through a two-layer FFN to obtain the output of the $k^{\text{th}}$ encoder stage:

$$X^{k+1}= X_{\text{MSA}}^{k} + E_2^k \Big( \text{ReLU} ( (E_1^k X_{\text{MSA}}^{k} + b_1^k) \Big) + b_2^k $$

\end{enumerate}

The output of the last encoder stage, $X^{K}$, will pass through a final FFN layer that projects it onto the ${\rm I\!R}^{N n_{\text{class}}}$ and then reshaped into ${\rm I\!R}^{n \times n \times n \times n_{\text{class}}}$, where $n_{\text{class}}$ is the number of classes (for binary segmentation, $n_{\text{class}}=2$):

$$\hat{Y}= \text{Softmax} \Big( E_{\text{out}} X^{K} + b_{\text{out}}) \Big). $$

\noindent $\hat{Y}$ is the predicted segmentation for the center patch of the block.

\subsection{Implementation and training}
\label{training}

The network was implemented in TensorFlow 1.16 and run on an NVIDIA GeForce GTX 1080 GPU on a Linux machine with 120 GB of memory and 16 CPU cores. We compare our model with 3D UNet++ \cite{zhou2018unet}, which is a state of the art CNN for medical image segmentation. 

We trained the network parameters to maximize the Dice similarity coefficient, DSC \cite{milletari2016}, between $\hat{Y}$ and the ground-truth segmentation of the center patch using Adam \cite{kingma2014}. We used a batch size of 10 and a learning rate of $10^{-4}$, which was reduced by half if after a training epoch the validation loss did not decrease.

\textbf{Pre-training:} Manual segmentation of complex structures such as the brain cortical plate can take several hours of a medical expert's time for a single image. Therefore, methods that can achieve high accuracy with fewer labeled training images are highly advantageous. To improve the network's performance when labeled training images are insufficient, we propose to first pre-train the network on unlabeled data for denoising or inpainting tasks. For denoising pre-training, we add Gaussian noise with $\text{SNR}= 10 \text{ dB}$ to the input block $B$, whereas for inpainting pre-training we set the center patch of the block to constant 0. In both cases, the target is the noise-free center patch of the block. For pre-training, we use a different output layer (without the softmax operation) and train the network to minimize the $\ell_2$ norm between the prediction and the target. To fine-tune the pre-trained network for the segmentation task, we introduce a new output layer with softmax and train the network on the labeled data as explained in the above paragraph. We fine-tune the entire network, rather than only the output layer, on the labeled images because We have found that fine-tuning the entire network for the segmentation task leads to better results.

\subsection{Data}

Table \ref{table:data} shows the datasets used in this work. We used $\sim1/5$ of the training images for validation. The final models were evaluated on the test images. The images were split into train/validation/test at random.

\begin{table*}[!htb]
 \caption{Datasets used for experiments in this work.}
  \label{table:data}
   \begin{center}
    \begin{tabular}{ L{3.6cm}  C{3.0cm} C{2.8cm} }
\hline
target organ & image modality & $[n_{train}, n_{test}]$   \\ \hline
Brain cortical plate & T2 MRI & [18, 9]  \\
Pancreas & CT & [231, 50]  \\
Hippocampus & MRI & [220, 40]  \\
\hline
\end{tabular}
  \end{center}
\end{table*}

\section{Results and Discussion}

\begin{figure}
\includegraphics[width=8.8cm]{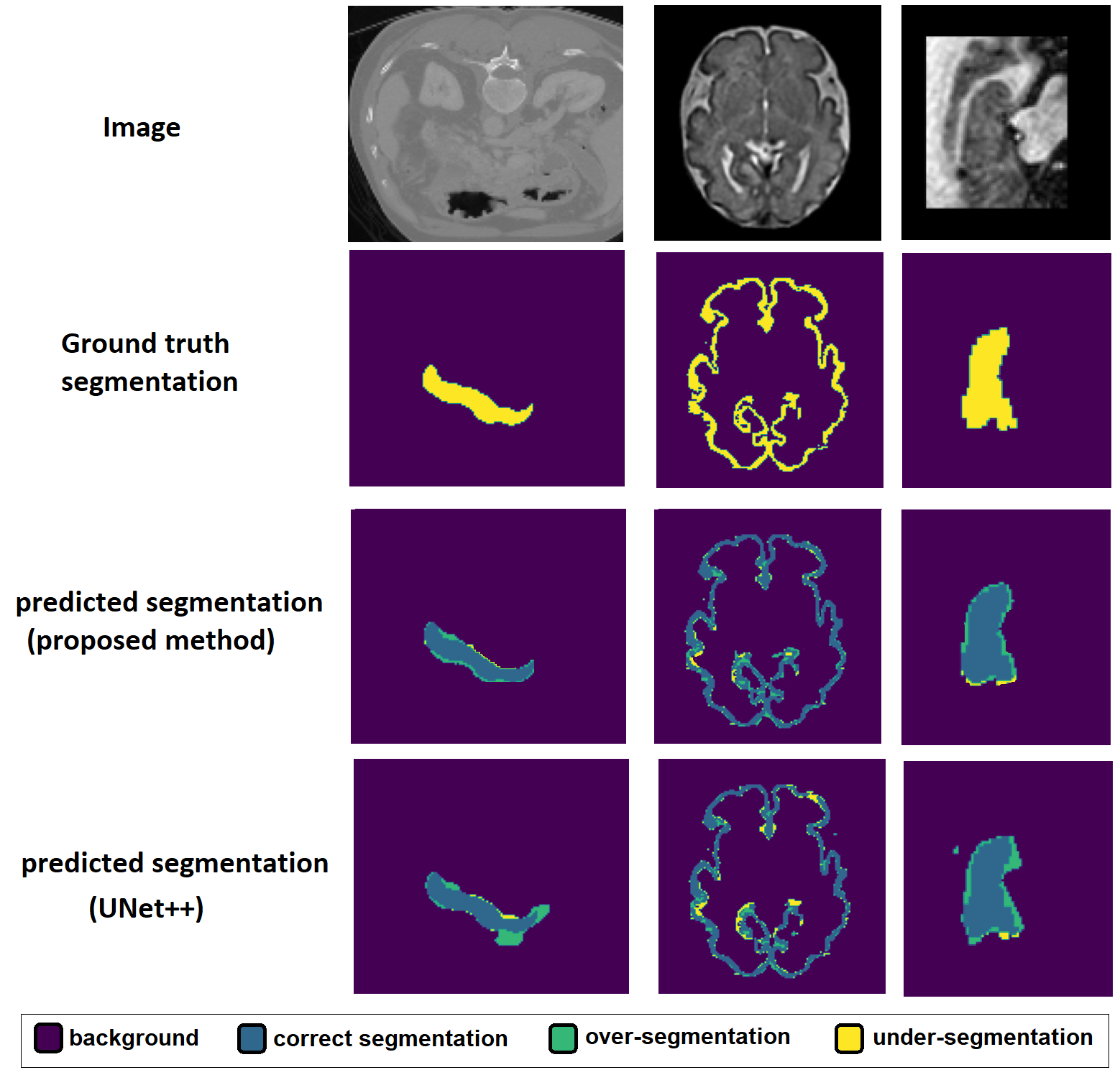}
\caption{Example segmentations predicted by the proposed method and UNet++.} 
\label{results_vis}
\end{figure}

Table \ref{table:results} presents test segmentation accuracy of the propsoed method and UNet++ in terms of DSC, the 95 percentile of the Hausdorff Distance (HD95) and Average Symmetric Surface Distance (ASSD). In these experiment, we used these parameter settings for our network: $K=7, W= 24, n=3, D= 1024, D_h= 256, n_h= 4$. We used the same parameter settings for all experiments reported in the rest of the paper, unless otherwise stated. For each dataset and each of the three criteria, we ran paired t-tests to see if the differences were statistically significant. As shown in the table, segmentation accuracy for the proposed convolution-free method was significantly better than, or at least on part with, UNet++. Figure \ref{results_vis} shows example slices from test images in each dataset and the segmentations predicted by the proposed method and UNet++. We have observed that the propsoed method is capable of accurately segmenting fine and complex structures such as the brain cortical plate. In terms of training time, our network converged in approximately 24 hours of GPU time.

\begin{table*}[!htb]
\footnotesize
 \caption{\footnotesize{Segmentation accuracy of the proposed method and UNet++. Better results for each dataset/criterion are in bold type. Asterisks denote statistically significant difference ($p<0.01$ in paired t-test.)}}
  \label{table:results}
   \begin{center}
    \begin{tabular}{ L{3.0cm} L{2.0cm} C{2.3cm} C{2.0cm} C{2.0cm} }
\hline
Dataset & Method & DSC & HD95 (mm) & ASSD (mm)   \\ \hline
\multirow{2}{*}{\parbox{3.0cm}{Brain cortical plate }} & Proposed & $\bm{0.879 \pm 0.026^*}$ & $0.92 \pm 0.04$ & $\bm{0.24 \pm 0.03}$  \\
& UNet++ & $0.860 \pm 0.024$ & $\bm{0.91 \pm 0.04}$ & $0.25 \pm 0.04$  \\ \hline
 \multirow{2}{*}{\parbox{3.0cm}{Pancreas }} & Proposed  & $\bm{0.826 \pm 0.020^*}$ & $\bm{5.72 \pm 1.61^*}$ & $\bm{2.13 \pm 0.24^*}$  \\
& UNet++  & $0.808 \pm 0.021$ & $6.67 \pm 1.80$ & $2.45 \pm 0.21$  \\  \hline
 \multirow{2}{*}{\parbox{3.0cm}{Hippocampus }} & Proposed  & $\bm{0.881 \pm 0.017^*}$ & $\bm{1.10 \pm 0.19}^*$ & $\bm{0.40 \pm 0.04}$  \\
& UNet++  & $0.852 \pm 0.022$ & $1.33 \pm 0.26$ & $0.41 \pm 0.07$  \\ 
\hline
\end{tabular}
  \end{center}
\end{table*}

\begin{figure*}
\centering
\includegraphics[width=184mm]{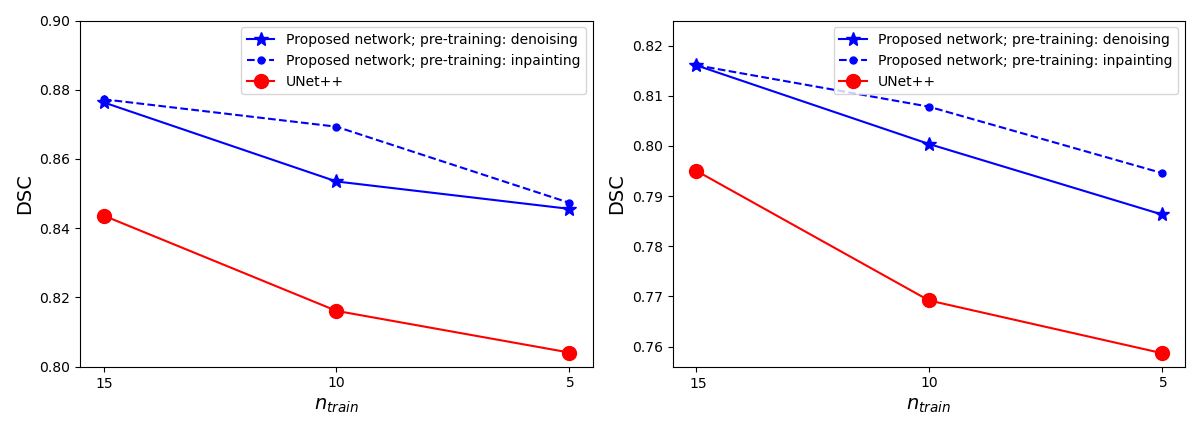}
\caption{Segmentation accuracy (in DSC) for the proposed network and UNet++ with reduced labeled training data on the cortical plate (left) and pancreas (right) datasets.} 
\label{results_downsample}
\end{figure*}

To assess the segmentation accuracy with reduced number of labeled training images, we trained our method and UNet++ with $n_{\text{train}}= 5, 10, \text{ and }15$ labeled training images from cortical plate and pancreas datasets. For cortical plate segmentation, we used 500 unlabeled images from the dHCP dataset \cite{bastiani2019} for pre-training. For pancreas segmentation, we used the remaining training images (i.e., $231-n_{\text{train}}$) for pre-training. We pre-trained our model as explained in Section \ref{training}. To pre-train UNet++, we used the method propsoed in \cite{bai2017semi}. Figure \ref{results_downsample} shows the results of this experiment. With the propsoed pre-training, the convolution-free network achieves significantly more accurate segmentations with fewer labeled training images.

\begin{figure}
\centering
\includegraphics[width=8.8cm]{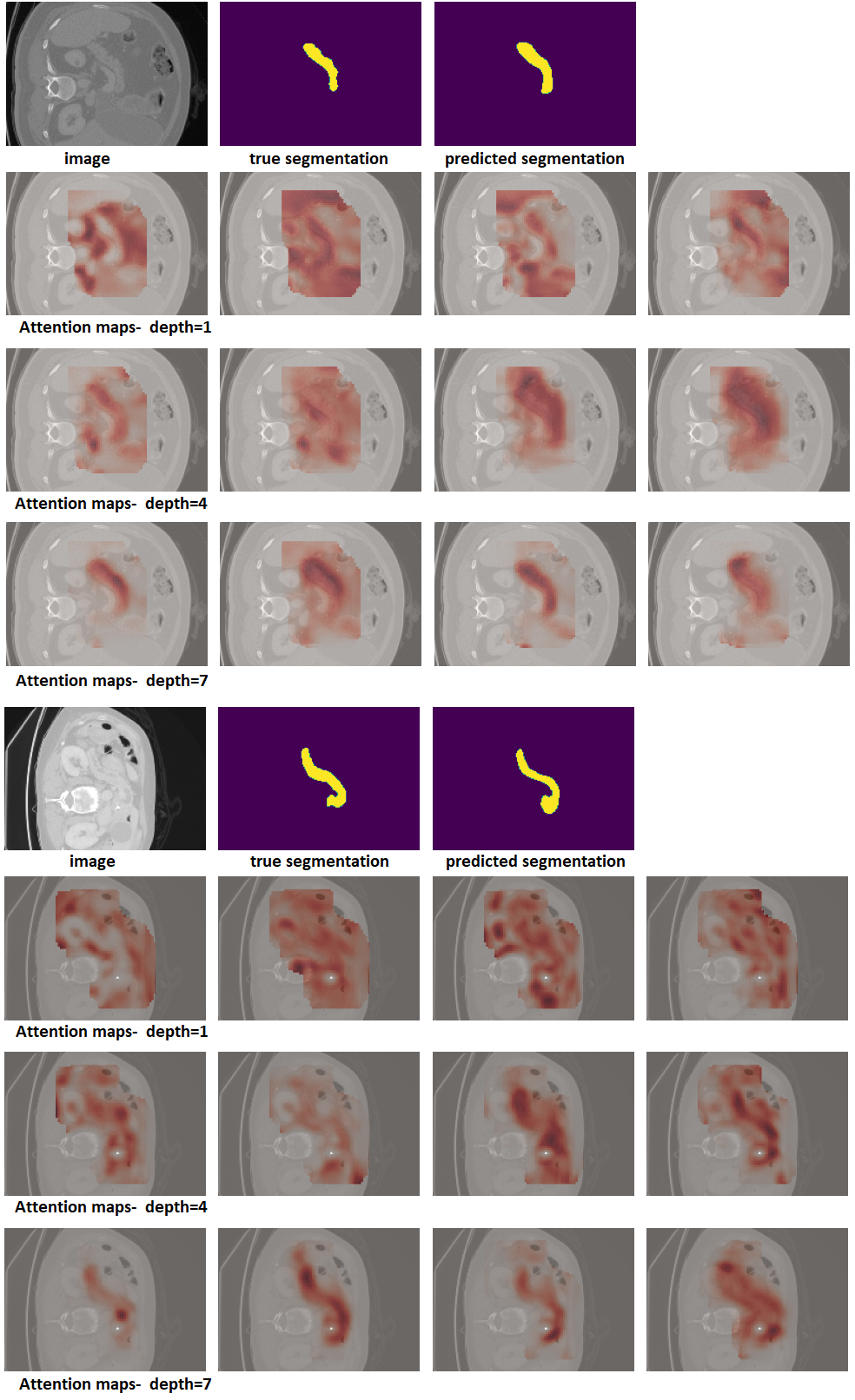}
\caption{Example attention maps for two pancreas images.} 
\label{attention_map_pancreas}
\end{figure}

Figure \ref{attention_map_pancreas} shows example attention maps of the proposed network for pancreas segmentation. As mentioned above, to process a test image of a desired size, we apply our network in a sliding window fashion. To generate the attention maps for the entire image, at each location of the sliding window the attention matrices (which are of size ${\rm I\!R}^{N \times N}$) are summed along the column to determine the total attention paid to each of the $N$ patches by the other patches in the block. Performing this computation in a sliding window fashion and computing the voxel-wise average gives us the attention maps shown in these figures. They indicate how much attention is paid to every part of the image.

\begin{figure}
\centering
\includegraphics[width=8.8cm]{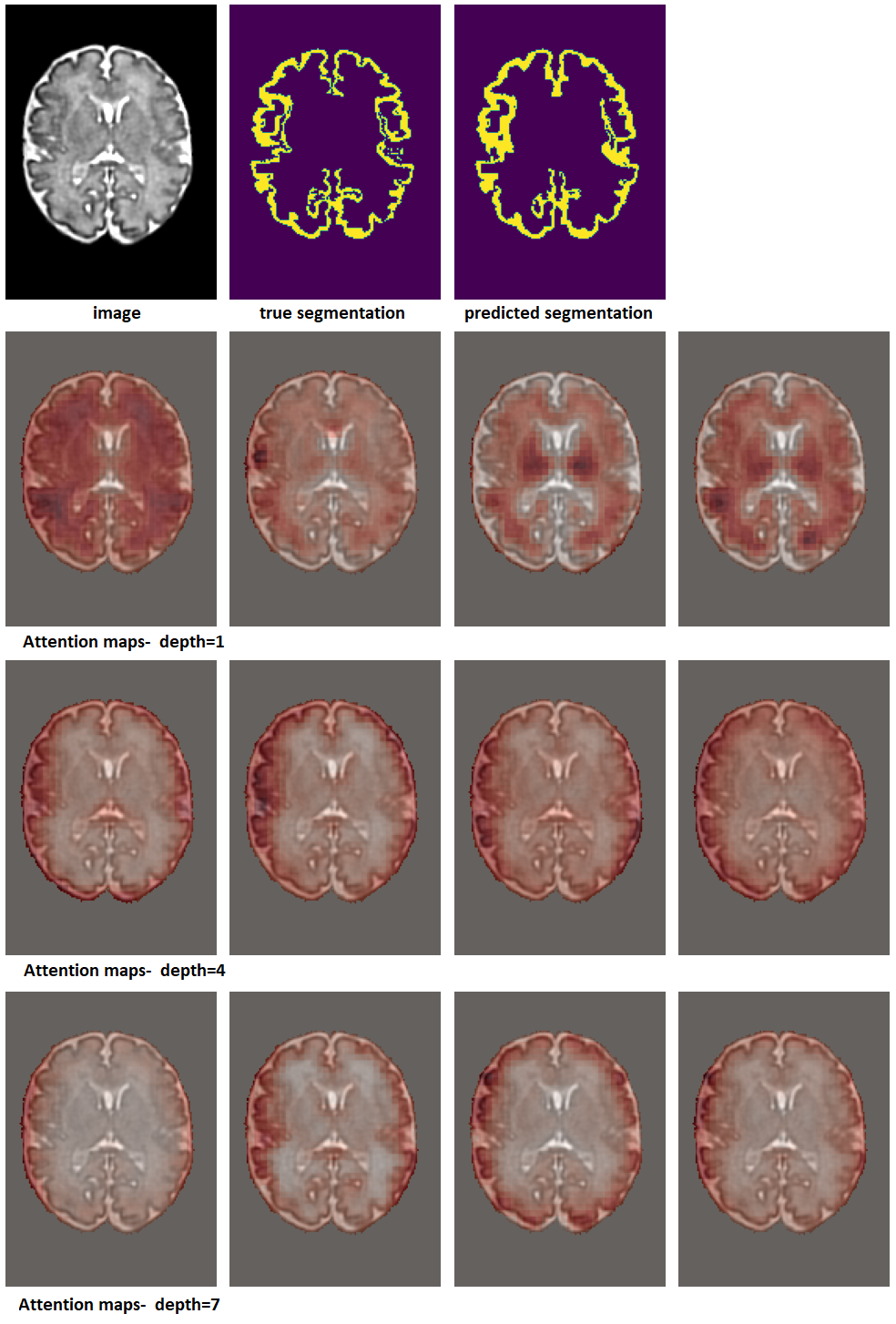}
\caption{Example attention maps for a cortical plate image.}
\label{attention_map_cp}
\end{figure}

The attention maps shown in Figure \ref{attention_map_pancreas} show that, in general, the early stages of the network have a wider attention scope. They attend to other structures and anatomical features that surround the organ of interest (here, the pancreas). The deeper stages of the network are more focused on the pancreas itself. A similar behaviour can be seen for cortical plate segmentation in Figure \ref{attention_map_cp}. In the earlier stage the network attends to the entire brain, while in the deeper layers the network's attention is more focused to the regions around the cortical plate. Another observation from these figures, especially Figure \ref{attention_map_pancreas} is the variability between the attention pattern in different heads of the multi-head self-attention mechanism. In each stage, the four separate heads of the MSA module adopt quite different attention patterns. This indicates that the multi-head design of the MSA gives the network more flexibility to learn the attention patterns that help improve the segmentation accuracy. The importance of multi-head design is well documented in natural language processing applications \cite{vaswani2017attention} and our results show it is important for 3D medical image segmentation as well, as we show further below.

Table \ref{table:parameters} effects of some of the network design choices on the segmentation accuracy on the pancreas dataset. In this table, the baseline (first row) corresponds to the settings that we have used in the experiments reported above, i.e., $K=7, W= 24, n=3, D= 1024, D_h= 256, n_h= 4$. We selected these settings based on preliminary experiments and we have found them to be good settings for different datasets. Increasing the number/size of the patches or increasing the network depth typically leads to slight improvements in accuracy. Furthermore, using a fixed positional encoding or no positional encoding slightly reduces segmentation accuracy compared with free-parameter/learnable positional encoding. Finally, using a single-head attention significantly reduces the segmentation accuracy, which indicates the importance of the multi-head design to enable the network to learn a more complex relation between neighboring patches.

\begin{table*}[!htb]
 \caption{\footnotesize{Effect of some of the network hyperparameters on the segmentation accuracy on the pancreas dataset. The baseline (first row) corresponds to these settings: $K=7, W= 24, n=3, D= 1024, D_h= 256, n_h= 4$, which are the hyperparameter values used in all experiments reported in this paper other than in this table. }}
  \label{table:parameters}
   \begin{center}
    \begin{tabular}{ L{4.0cm} C{2.5cm} C{2.0cm} C{2.0cm} }
\hline
Parameter settings & DSC & HD95 (mm) & ASSD (mm)   \\ \hline
baseline  & $0.826 \pm 0.020$ & $5.72 \pm 1.61$ & $2.13 \pm 0.24$  \\
larger blocks, $n=5$  & $0.828 \pm 0.023$ & $5.68 \pm 1.63$ & $2.01 \pm 0.18$  \\
no positional encoding  & $0.818 \pm 0.026$ & $5.84 \pm 1.74$ & $2.26 \pm 0.27$  \\
fixed positional encoding  & $0.823 \pm 0.021$ & $5.81 \pm 1.54$ & $2.16 \pm 0.23$  \\
deeper network, $K=10$  & $0.827 \pm 0.018$ & $5.50 \pm 1.48$ & $2.10 \pm 0.21$  \\
shallower network, $K=4$    & $0.810 \pm 0.023$ & $6.14 \pm 1.80$ & $2.29 \pm 0.38$  \\
more heads, $n_h= 8$    & $0.824 \pm 0.017$ & $5.68 \pm 1.56$ & $2.14 \pm 0.21$  \\
single head, $n_h= 1$   & $0.802 \pm 0.026$ & $6.82 \pm 1.40$ & $2.31 \pm 0.35$  \\
\hline
\end{tabular}
  \end{center}
\end{table*}

\section{Conclusions}

The convolution operation has a strong basis in the structure of the mammalian primary visual cortex and it is well suited for developing powerful techniques for image modeling and image understanding. In recent years, CNNs have been shown to be highly effective in tackling various computer vision problems. However, there is no reason to believe that no other model can outperform CNNs on a specific vision task. Medical image analysis applications, in particular, pose specific challenges such as 3D nature of the images and small number of labeled images. In such applications, other models can be more effective than CNNs. In this work we presented a new model for 3D medical image segmentation. Unlike all recent models that use convolutions as their main building block, our model is based on self-attention between neighboring 3D patches. Our results show that the proposed network can outperform a state of the art CNN on three medical image segmentation datasets. With pre-training for denoising and in-painting tasks on unlabeled images, our network also performed better than a CNN when only 5-15 labeled training images were available. We expect that the network proposed in this paper should be effective for other tasks in medical image analysis such as anomaly detection and classification.

\vspace{2mm}

\section*{Acknowledgment}
\noindent This project was supported in part by the National Institute of Biomedical Imaging and Bioengineering and the National Institute of Neurological Disorders and Stroke of the National Institutes of Health (NIH) under award numbers R01EB031849, R01NS106030, and R01EB032366; in part by the Office of the Director of the NIH under award number S10OD0250111; in part by the National Science Foundation (NSF) under award 2123061; and in part by a Technological Innovations in Neuroscience Award from the McKnight Foundation. The content of this paper is solely the responsibility of the authors and does not necessarily represent the official views of the NIH, NSF, or the McKnight Foundation.

The DHCP dataset is provided by the developing Human Connectome Project, KCL-Imperial-Oxford Consortium funded by the European Research Council under the European Union Seventh Framework Programme (FP/2007-2013) / ERC Grant Agreement no. [319456]. We are grateful to the families who generously supported this trial.

\bibliographystyle{splncs04}
\bibliography{davoodreferences}

\end{document}